\documentclass{aastex}

\usepackage{amsmath}
\usepackage{amssymb}
\usepackage{latexsym}
\usepackage{graphicx}
\usepackage{natbib}
\bibpunct{(}{)}{;}{a}{}{,}

\begin{document}
\title{High-resolution observations of active region moss and its dynamics}

\author{R. J. Morton and J. A. McLaughlin} 
\affil{Department of Mathematics \& Information Sciences, Northumbria University, Newcastle Upon Tyne,
NE1 8ST, UK}
\email{richard.morton@northumbria.ac.uk}

\begin{abstract}
The \textit{High resolution Coronal Imager (Hi-C)} has provided the sharpest view of the EUV corona to date. In this paper we exploit its impressive
resolving power to provide the first analysis of the fine-scale structure of moss in an active region. The data reveal that the moss is made up
of a collection of fine threads, that have widths with a mean and standard deviation of $440\pm190$~km (Full Width Half Maximum). {The 
brightest moss emission is located at the visible head of the fine-scale structure and the fine structure appears to extend into the lower solar 
atmosphere.} The emission decreases along the features implying the lower sections are most likely dominated by 
cooler transition region plasma. These threads appear to be the cool, lower legs of the hot loops.
In addition, the increased resolution allows for the first direct observation {of physical displacements of the moss fine-structure in a direction transverse 
to its central axis. Some of these transverse displacements demonstrate periodic behaviour, which we interpret as a signature of    
kink (Alfv\'enic) waves. Measurements of the properties of the transverse motions are made and the wave motions have} means and 
standard deviations of $55\pm37$~km for the transverse displacement amplitude, $77\pm33$~s for the period and $4.7\pm2.5$~km/s for the velocity 
amplitude. The presence of waves in the transition region of hot loops could have important implications 
for the heating of active regions.
\end{abstract}


\date{}

\shorttitle{High-resolution observations of the dynamic moss}

\shortauthors{Morton and McLaughlin}
\maketitle

\section{Introduction}
One of the fundamental and persistent problems in astrophysics is the puzzle of how the solar corona is heated. There has been a wide range of 
scenarios proposed to explain the observed high temperatures ($T>1$~MK), e.g., magnetic reconnection (nanoflares - \citealp{PAR1988}), 
magnetohydrodynamic (MHD) waves (\citealp{CRAetal2007}) and type-II spicules (\citealp{DEPetal2011}) - although it is not necessary that each of these 
are exclusive. It is generally accepted that the contributing processes are likely to occur on small spatial and temporal scales.

The \textit{High resolution Coronal Imager (Hi-C)} (\citealp{KOBetal2014}) provided a unique view of the EUV corona and, despite the relatively 
short lifetime of the mission ($<300$~s), has allowed for a number of insights into small-scale coronal features (\citealp{BROetal2013}; 
\citealp{PETetal2013}; \citealp{ALEetal2013}). In particular, \textit{Hi-C} allowed for a detailed study of the 
moss regions (\citealp{TESetal2013}; \citealp{WINetal2013}), i.e., the upper Transition Region emission of high pressure loops 
in active regions. The moss appears as a reticulated pattern of bright emission in EUV images, with large-scale structuring on spatial scales 
of 2-3~Mm, with an apparent vertical extent of  $\sim1-4$~Mm (\citealp{BERetal1999}; \citealp{FLEDEP1999}). The bright emission is punctuated with 
patches of low emission ({\lq dark inclusions\rq}) and, in general, the regions of low emission show a correlation with spicules observed in 
H$\alpha$ wings (\citealp{DEPetal2003}). However, previous instruments have not had the ability to resolve fine-scale structure in either the bright or 
dark regions.

Moss has been the focus of much interest (e.g., \citealp{TRIetal2010}; \citealp{BROetal2010}) since it was proposed that the moss emission
scales well with loop pressure (\citealp{MARetal2000}), enabling variations in the moss to provide a diagnostic tool for the study of coronal 
heating mechanisms. Moreover, observations have revealed that moss emission varies little over extended time periods (e.g., \citealp{ANTetal2003}; 
\citealp{BROWAR2009}), implying that the heating must be quasi-steady in nature and dominated by continuous high-frequency heating 
events. This scenario has support from reports of high-frequency intensity variations in the moss, observed with the high spatial and temporal resolution 
of \textit{Hi-C} (\citealp{TESetal2013}). On the other hand, time variability of the moss could well be due to motions of the magnetic field rather than a 
direct signature of heating. {This was suggested by \cite{ANTetal2003} and \cite{BROWAR2009} in relation to variability on long time-scales, but 
could also apply to the variability observed on shorter time-scales.}

In recent years, the role of MHD waves in heating has been brought to the forefront of the field due to observations of ubiquitous 
kink (Alfv\'enic) wave behaviour throughout the chromosphere (\citealp{DEPetal2007}; \citealp{KURetal2012}; \citealp{MORetal2012c, MORetal2013}) 
and corona (\citealp{TOMetal2007}; \citealp{VANetal2008}; \citealp{ERDTAR2008}; \citealp{MCIetal2011}). In particular, the observed
chromospheric waves have an estimated wave energy flux in excess of that needed to meet the heating requirements of the active corona.
However, current observations of EUV coronal loops reveal that the kink wave energy flux in the corona is generally too small to contribute to heating in 
the coronal volume (\citealp{TOMetal2007}; \citealp{MCIetal2011}; \citealp{MORMCL2013}). This lack of observed wave energy may be in part due to 
wave reflection at the Transition Region (e.g., \citealp{OKADEP2011}) or the waves may have been significantly damped/undergone mode
conversion before reaching the observable EUV corona (e.g., \citealp{VERTHetal2010}; \citealp{MORetal2013b}). To date, it has not been possible to carry 
out similar wave studies for warm/hot loops due to a combination of low spatial resolution, low signal-to-noise and the increased {\lq fuzziness\rq} of 
warm loops in imaging observations (e.g., \citealp{BRISCH2006}). 

\textit{Hi-C} has provided images of resolved fine-scale structure in coronal loops and while studies 
have exploited the high spatial and temporal resolution of \textit{Hi-C} to investigate the temporal variations in moss regions, the spatial structure has 
not yet been examined. Here we provide the first analysis of the fine-scale spatial structure of the moss regions. {It is found that the bright moss 
is located at the upper end of elongated fine-structure} that has spatial scales of a few 100~km, similar to those observed in the chromospheric fine-
structure (e.g., \citealp{MORetal2012c}; \citealp{ANTROU2012}; \citealp{PERetal2012}) and EUV loops (e.g., 
\citealp{BROetal2012, BROetal2013}; \citealp{PETetal2013}). The fine-structure appears to be the lower (upper chromosphere/lower Transition Region) 
legs of the hot loops typically seen in soft X-rays. The ability to resolve the fine-scale structure associated with the moss also allows for the first imaging 
observations of {transverse displacements of the structures}. In particular,  periodic transverse displacement of the fine-scale structure is 
observed and interpreted in terms of the kink (Alfv\'enic) mode. Measurements demonstrate the waves have periods and amplitudes similar to or smaller 
than those found previously in fibrils and spicules, respectively (e.g., \citealp{OKADEP2011}; \citealp{PERetal2012}; \citealp{KURetal2012}; 
\citealp{MORetal2012c, MORetal2013}). 

\begin{figure*}[!tp]
\centering
\includegraphics[scale=0.79, clip=true, viewport=1.8cm 1.0cm 11.cm 10.6cm]{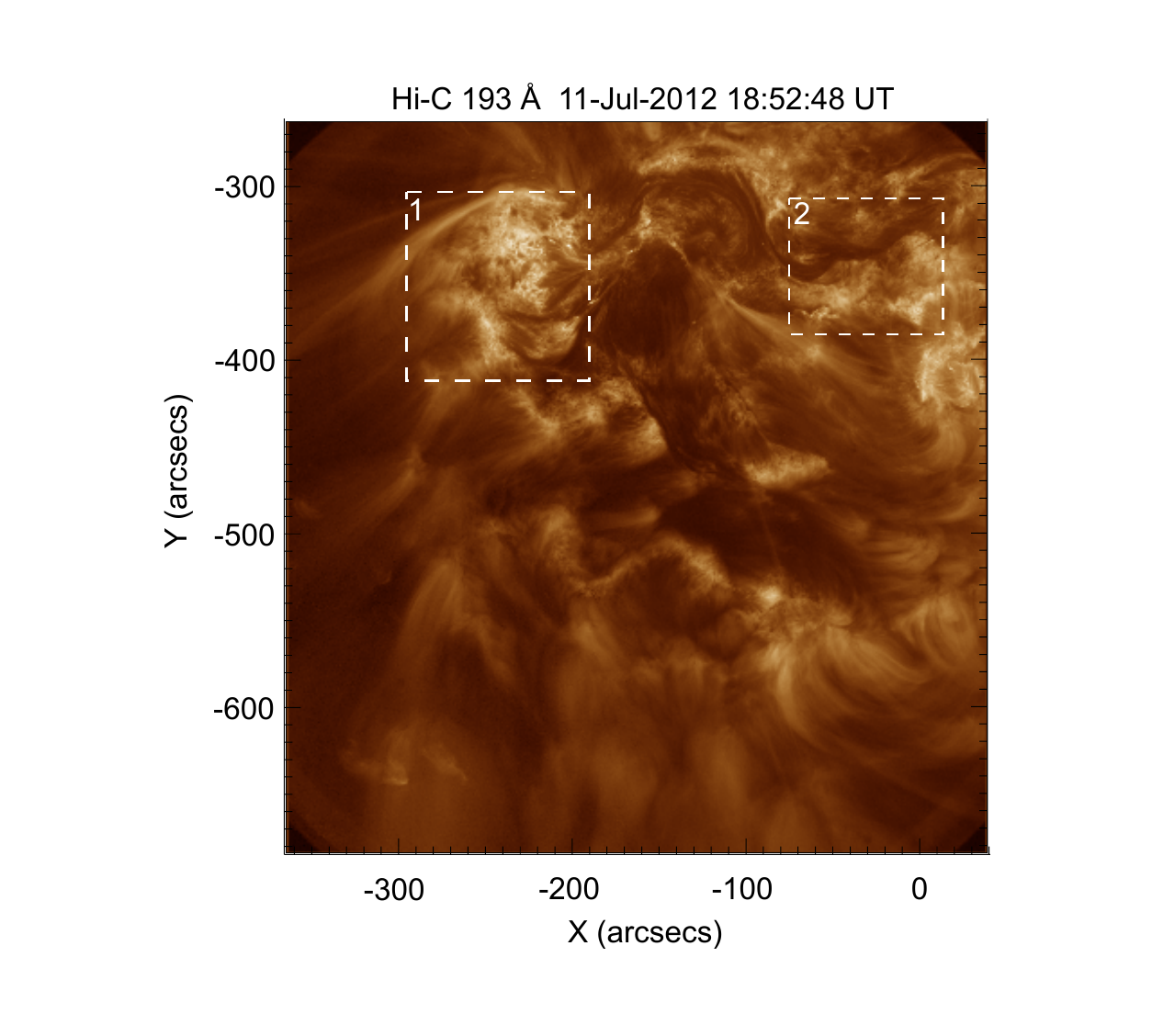} 
\includegraphics[scale=0.33, clip=true, viewport=0.0cm -0.5cm 25.cm 23.cm]{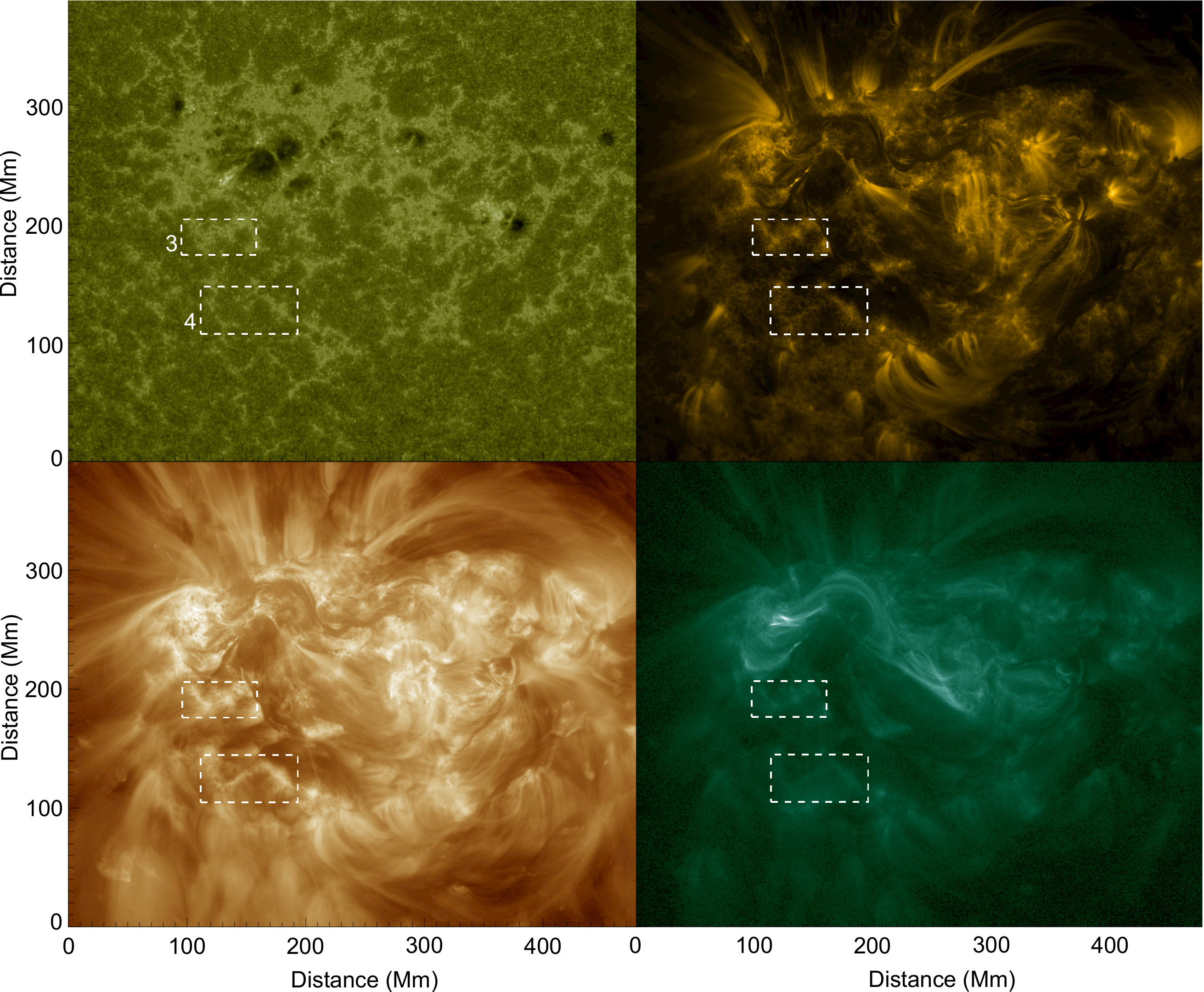} 
\caption{Left panel: Active region observed with \textit{Hi-C} $193$~{\AA}. The moss regions at the foot-points of hottest coronal loops in the active
region are indicated by the dashed boxes and are arbitrarily labelled 1 and 2. Right panel: The same active region observed with \textit{AIA} 1600~{\AA} (top left), 171~{\AA} (top right), 193~{\AA} (bottom left) and 94~{\AA} (bottom right). The boxed areas show two additional regions considered to be moss and are arbitrarily labelled 3 and 4.}\label{fig:fov}
\end{figure*}

\section{Observations and data reduction}
Details of the \textit{Hi-C} observations can be found in, e.g., \cite{MORMCL2013}. We note here that the cadence of the data is on average 5.57~s, a 
correction from the cadence given in \cite{MORMCL2013}. The data was 
processed and aligned by the \textit{Hi-C} science team, however, we note that 
the data still displayed visible shifts and additional alignment was performed using 
cross-correlation, achieving sub-pixel accuracy on the frame-to-frame alignment. The data are missing a frame between 
{18:54:28-18:54:40}~UT (time stamp corrected), therefore we used interpolation to create the missing frame and 
provide a constant sampling rate for wave studies. {The first seven frames of the \textit{Hi-C} data are subject to viewing distortions due to rocket 
jitter and prove inadequate for rigid alignment; hence, they are not used for analysis.} In addition,
we compare the \textit{Hi-C} data to data from the \textit{Solar Dynamic Observatory (SDO) Atmospheric Imaging Assembly (AIA)} 
(\citealp{LEMetal2011}), which is prepared using the standard 
routines. Alignment between \textit{Hi-C} and \textit{SDO} images is performed by degrading  the spatial sampling of the \textit{Hi-C} data to match 
that of \textit{SDO} and using cross-correlation.

\begin{figure*}[!htp]
\centering
\includegraphics[scale=0.6, clip=true, viewport=0.0cm 0.0cm 26.5cm 19.cm]{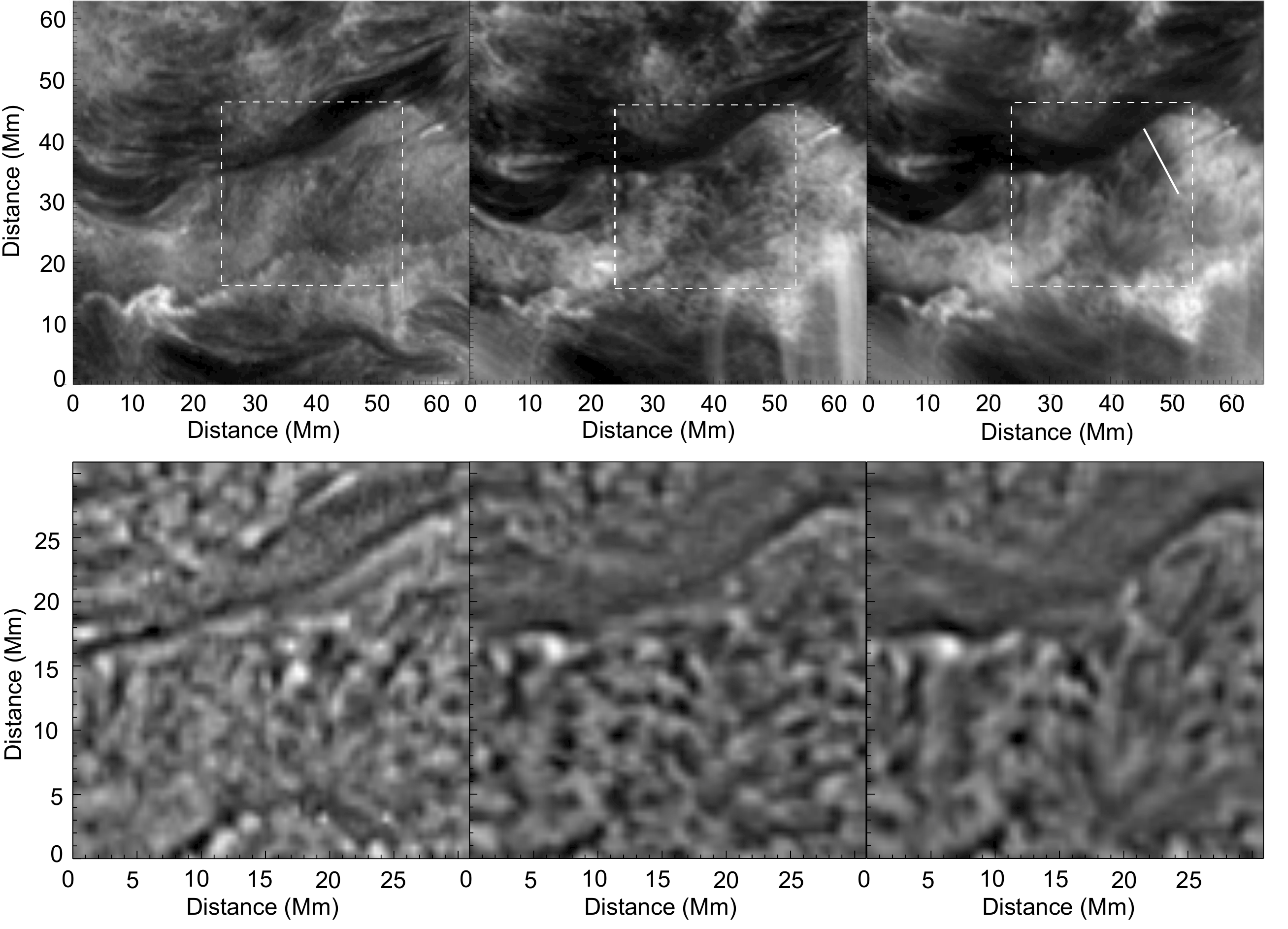} 
\caption{Top row: Images from left to right are \textit{AIA} 304~{\AA}, 171~{\AA} and 193~{\AA} of region 2. The bright emission 
corresponding to the moss regions is clearly visible in 171~{\AA} and 193~{\AA}, along with the dark inclusions. The solid white line indicates the cross-cut position used in Figure~\ref{fig:mossa}. Bottom row: Unsharp masked images 
of the dashed boxed regions shown in the top row. 
}\label{fig:moss}
\end{figure*}

The \textit{Hi-C} bandpass is centred close to 193~{\AA}, which has strong contributions from \ion{Fe}{12} that has a peak formation temperature close 
to $1.5$~MK and is ideal for observing coronal features. As reported in \cite{MORMCL2013} and \cite{BROetal2013}, the images show apparently 
resolved coronal loops (Figure~\ref{fig:fov}). In addition, it is evident that there exists fine structure in the moss regions. It is these features that are the 
subject of the following investigation.  

Figure~\ref{fig:fov} shows the \textit{Hi-C} field of view and the brightest moss regions are indicated by the dashed boxes. \cite{TESetal2013} 
demonstrated that these regions lie at the foot-points of the hottest coronal loops in the active region,
which have significant X-ray emission observed in co-temporal observations from \textit{Hinode X-Ray Telescope (XRT} - \citealp{GOLetal2007}). These 
hot loops are also observed in Figure~\ref{fig:fov} in 94~{\AA} (\ion{Fe}{18}). While these are likely the hottest loops in the region, we argue that 
there are other hot (or at least warm $T>2$~MK) loops in the \textit{Hi-C} field of view. In Figure~\ref{fig:fov} we identify two additional regions 
that we suggest can be classified as moss. Firstly, bright patches in 1600~{\AA} (continuum plus \ion{C}{4}) images provide a good proxy for identifying 
magnetic elements and both the highlighted regions show enhanced, plage-like emission. Secondly, these regions in 171~{\AA} (\ion{Fe}{9}) images 
show the reticulated emission typically associated with moss, with no evident coronal loop structures
originating in the regions. This is in contrast to 193~{\AA}, which demonstrates the presence of very fine-scale, diffuse loops that are apparently rooted 
in the identified regions. In the 94~{\AA} bandpass, these diffuse fine-scale loops appear as a haze of emission, 
with the emission above the bright {\lq moss\rq} having a marginally greater intensity. The fuzziness of the loops would suggest they are $T<3$~MK 
(\citealp{REALetal2011}), although the identification of any fine-structure in the 94~{\AA} channel is restricted due to the lower signal-to-noise, i.e., 
compared to the 193~{\AA} channel. In \textit{XRT} images, region 3 has faint emission (compared to the hot loops) while region 4 lies outside the \textit{XRT} field of view (see Figure~1 of \citealp{TESetal2013}).

Taking into account the thermal responses of the \textit{AIA} channels, the lack of emission for the diffuse fine-scale loops in the 171~{\AA} channel and 
the presence of emission in 193~{\AA} and 94~{\AA} channels, this suggest the moss regions are foot-points of warm or hot loops ($T>2$~MK). The 
temperature of these threads requires there to be enhanced pressure at the loop foot-points, which in turn leads to enhanced emission in the Transition 
Region. The weaker 
emission of the moss (relative to the
bright moss regions 1 and 2) suggests that the pressure at the Transition Region of these loops is lower or the loop filling factor is less than the hottest 
loops observed in 94~{\AA} (\citealp{MARetal2000}).

\begin{figure*}[!htp]
\centering
\includegraphics[scale=0.63, clip=true, viewport=0.0cm 0.0cm 27.cm 8.2cm]{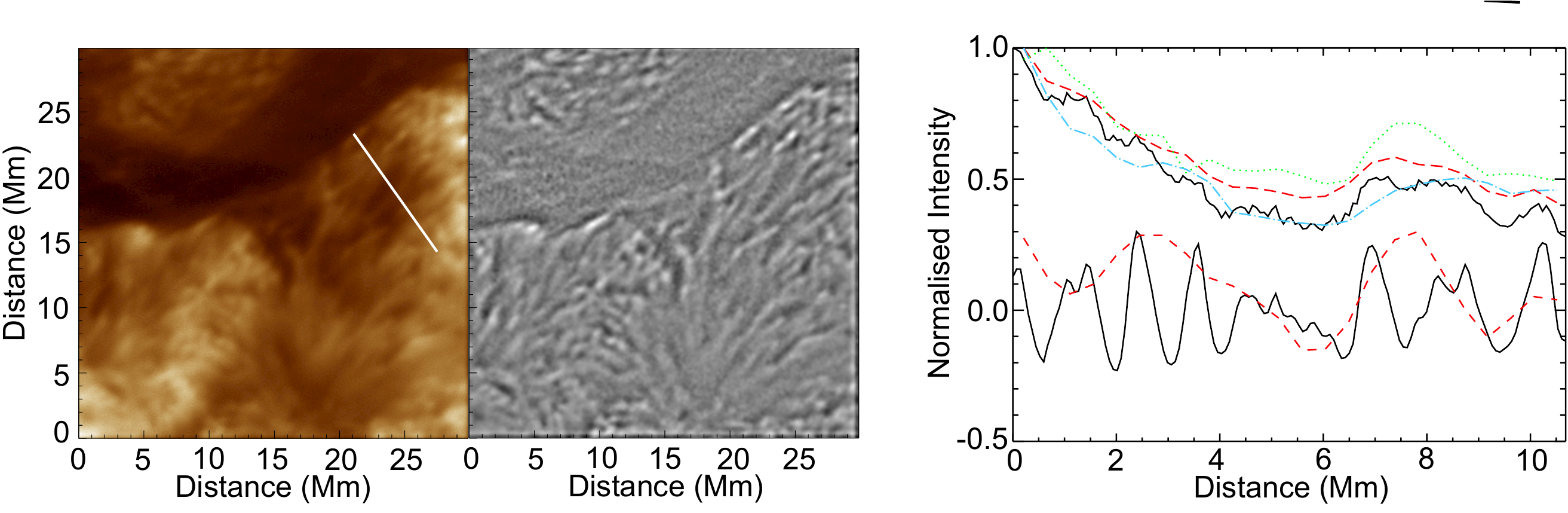} 
\caption{The left hand panels show a view of the 
moss region in the dashed box (Figure~\ref{fig:moss}) as observed with \textit{Hi-C} and an unsharped masked version of the region, which clearly 
reveals the fine-scale structure. The solid white line indicates the cross-cut position used in the far 
right hand panel. The separate right hand panel is 
the normalised intensities taken along co-spatial slits in \textit{Hi-C} (black) and \textit{AIA} (304~{\AA} - green/dotted, 171~{\AA} - blue/dash-dot, 
193~{\AA} - red/dash). The lower set of lines correspond to the same cross-cut from unsharp masked \textit{Hi-C} and \textit{AIA} 193~{\AA} images.
}\label{fig:mossa}
\end{figure*}

\begin{figure*}[!htp]
\centering
\includegraphics[scale=0.6, clip=true, viewport=0.0cm 0.0cm 27cm 12.cm]{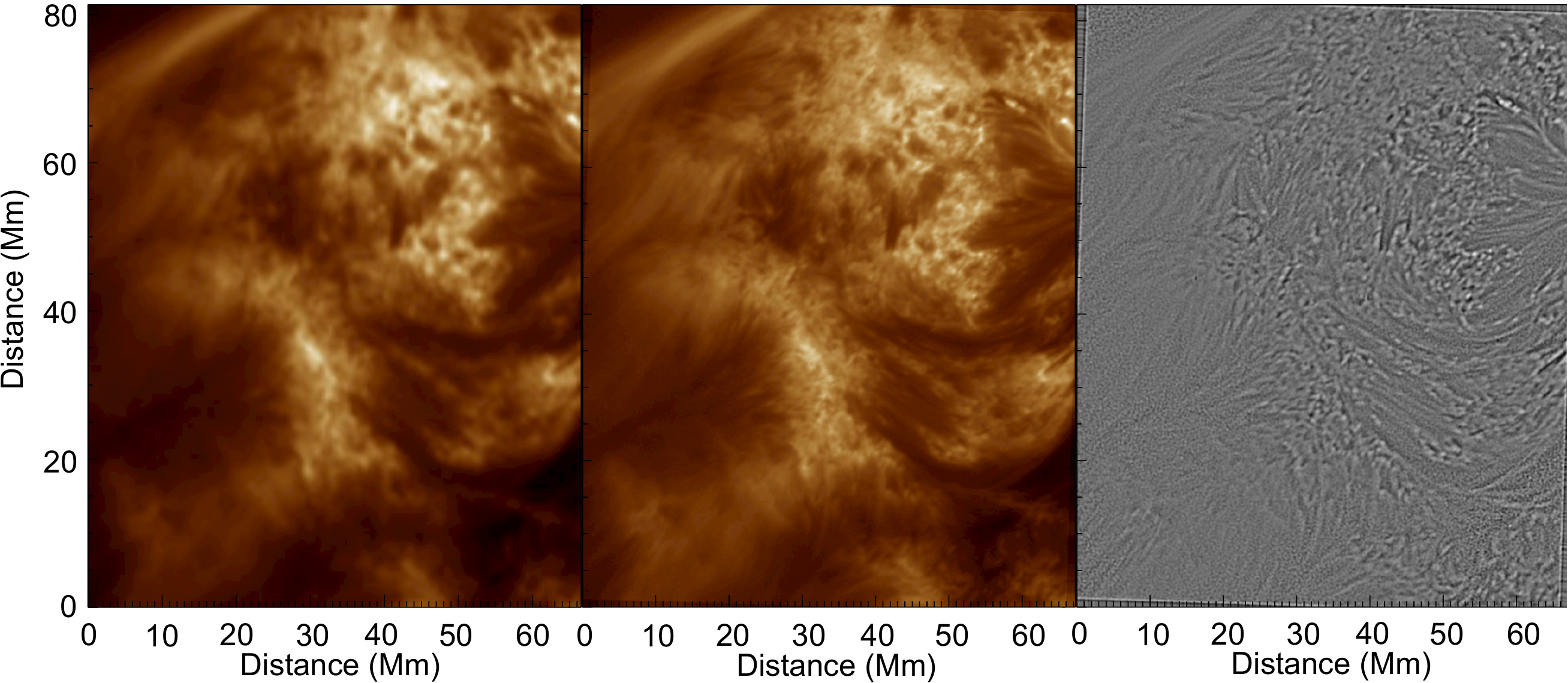} 
\caption{Close up images from the moss region labelled 1. From left to right, the panels show \textit{AIA} 193~{\AA}, \textit{Hi-C} and unsharp masked \textit{Hi-C} images, respectively. 
}\label{fig:moss2}
\end{figure*}

\begin{figure*}[!htp]
\centering
\includegraphics[scale=0.56, clip=true, viewport=1.0cm 1.5cm 26.cm 12.5cm]{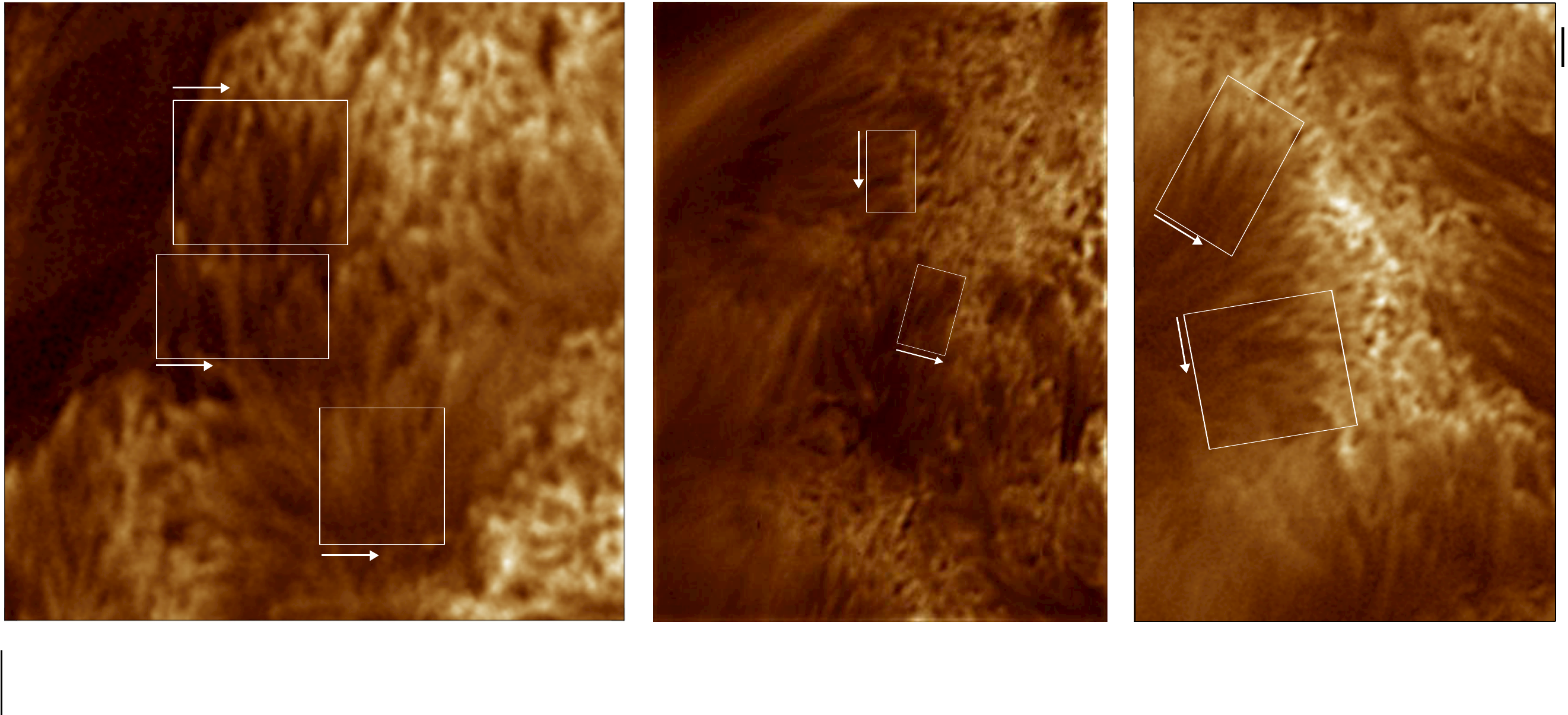} 
\caption{Variation along the moss fine-structure. Three enhanced images of moss regions are displayed and
white boxes highlight groups of fine structure. The intensity profiles are averaged across the direction indicated by the
white arrows, hence provide intensity profiles parallel to axis of the fine structure (see, Figure~\ref{fig:moss3}).
}\label{fig:moss2a}
\end{figure*}

\section{Results}
\subsection{Fine-scale structure of the moss}
We begin by providing a view of moss as it is observed with \textit{AIA}. Figure~\ref{fig:moss} (top row) shows a close up of region 2  in \textit{AIA} 
304~{\AA}, 171~{\AA} and 193~{\AA} bandpasses. The \textit{AIA} images suggest that small-scale structuring in the moss regions is present but it is 
clearly unresolved. In Figure~\ref{fig:mossa}, we display an image taken with \textit{Hi-C} that focuses on the same patch of the moss as shown in the 
bottom row of Figure~\ref{fig:moss}. The fine structuring is now evident appearing as threads, but the threads visibility is clearer after 
passing the data through an unsharp-mask routine. {The \textit{Hi-C} data reveals that the fine structures are connected to the bright moss and 
appear to be an extension of the bright moss into the lower solar atmosphere. }

\begin{figure}[!tp]
\centering
\includegraphics[scale=0.5, clip=true, viewport=0.0cm 0.0cm 17.5cm 12.5cm]{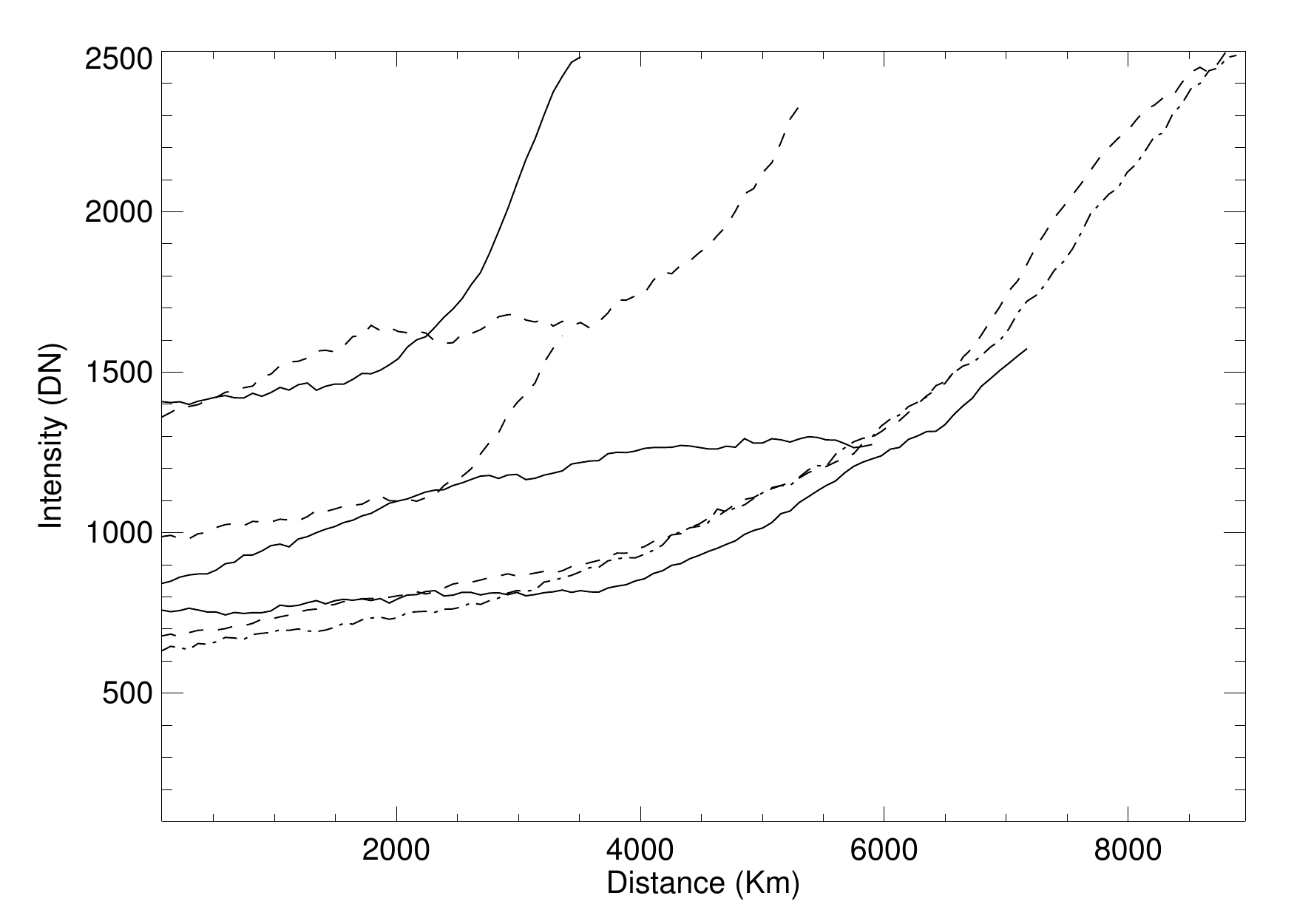} 
\caption{Intensity profiles along the moss fine-structure. The plot displays the average intensity profiles obtained from 
different groups of moss structures in regions 1 and 2. The intensity is measured from the visible footpoint of the fine-structure to the
bright moss at the head of the features.
}\label{fig:moss3}
\end{figure}

{These fine threads are visible in the dark inclusions, hence have reduced emission relative to the bright moss. Figure~\ref{fig:moss3} shows the 
intensity profiles of seven typical groups of the fine-structure (displayed in Figure~\ref{fig:moss2a}), where the profile is parallel to the axis and averaged 
over the group (groups consisting of 15-20 features). The intensity along the structures is found to steadily decrease from the bright moss into the dark inclusions.} Previous limb 
observations with \textit{TRACE} data demonstrated that moss has an apparent vertical extent on the order of $3000-4000$~km 
(\citealp{MARetal2000}). The fine-scale structures seen here appear to be have similar vertical scales (with some more extended moss regions), although 
it is difficult to locate where individual threads end/begin (see, Figures~\ref{fig:mossa} and \ref{fig:moss2}).

{The particular dark inclusion shown in Figure~\ref{fig:mossa} is a gap between two patches of bright moss. Each patch of moss can be seen to be 
composed of groups of fine-scale structure and the groups are inclined at different orientations. The variation in inclination of individual moss patches 
has been noted previously by \cite{KATTSU2005}.}

In Figure~\ref{fig:moss2} we show an example of a larger patch of moss corresponding to region 1. On viewing 
the unsharp masked \textit{Hi-C} image, the wealth of fine-scale structure in the moss regions is evident. The moss region centred at (30,30) provides a 
clear 
demonstration of the extension of the fine structure. The bright moss emission is at the head of the fine-scale structures, which extend off to 
the left hand side, gradually fading and becoming indistinguishable from the background emission. The formation of the structures is reminiscent of a 
chain of mountains, with the fine structure forming the sides of the mountain, meeting in the middle with the bright moss emission as the peaks. This 
formation is often seen in active regions imaged with H$\alpha$ (e.g., \citealp{DEPetal2003,DEPetal2007c}; \citealp{KURetal2011}), where the 
chromospheric structures tend towards a central location, occasionally having enhanced emission.

The difference between the \textit{Hi-C} and the \textit{AIA} view of the moss is elucidated in Figure~\ref{fig:mossa}. Taking a cross-cut perpendicular to 
the fine structure and plotting the intensity reveals that the individual strands are unresolved by \textit{AIA}, while \textit{Hi-C} shows peaks in 
emission related to the fine structure. Again, the fine structure is better visualised by comparing the intensity profiles from unsharp masked images (i.e. 
after subtracting the local mean intensity) for the same cross-cuts for \textit{Hi-C} and \textit{AIA}. 

\begin{figure}[!tp]
\centering
\includegraphics[scale=0.3, clip=true, viewport=0.0cm 0.0cm 26.5cm 19.3cm]{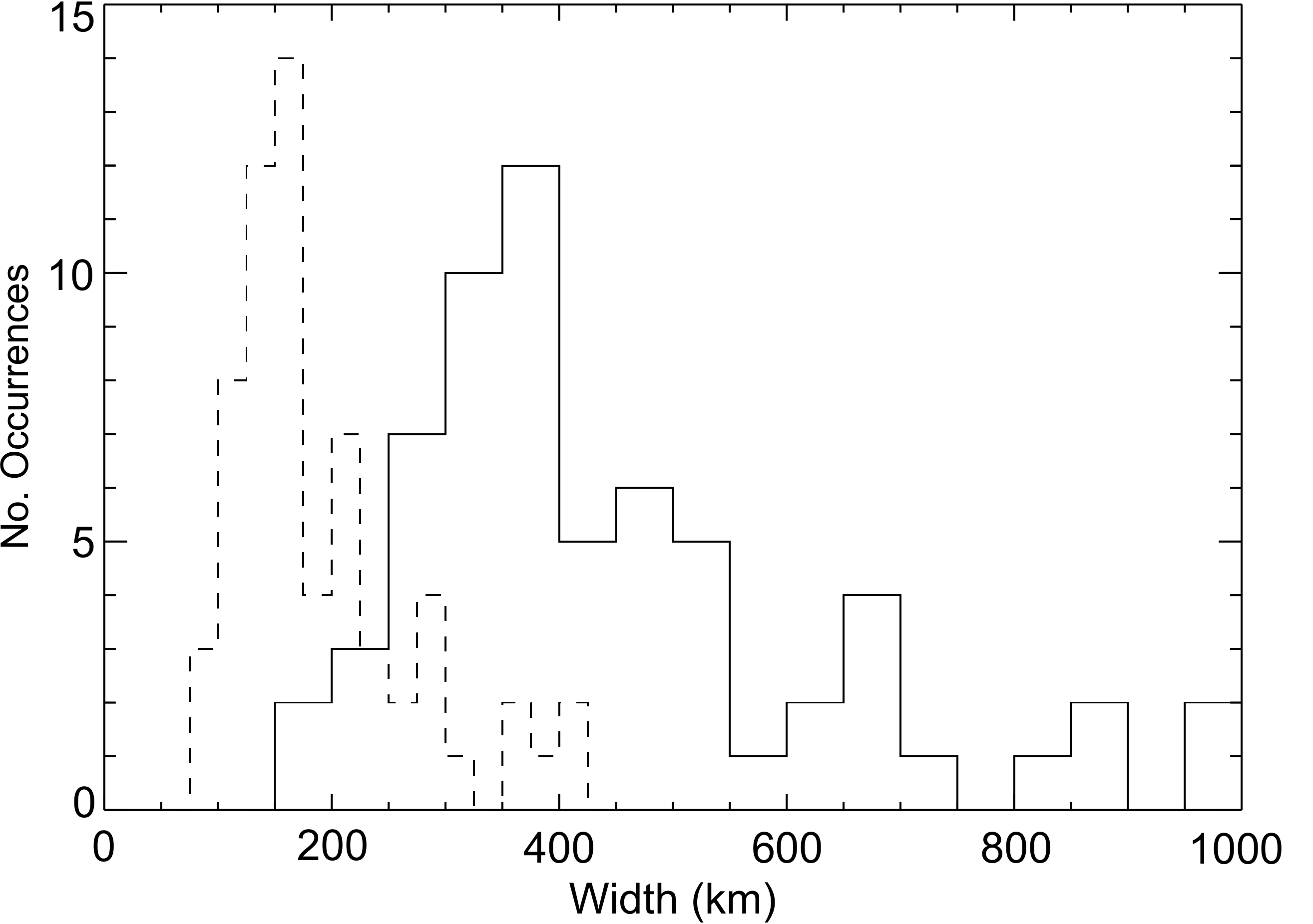} 
\caption{Histogram of measured widths for structures in the moss regions. The dashed lines show the $\sigma$ values for the fitted Gaussians
and the solid lines are the Full-Width-Half-Maximum values. The results have a mean width of $\sigma=188\pm79$~km, minimum $\sigma=76$~km 
and maximum $\sigma=414$~km.}\label{fig:moss_hist}
\end{figure}

In order to reveal the typical scale of the fine structures, we select features from each of the identified moss regions and measure their 
widths in the \textit{Hi-C} data. The fine structure in unsharp masked images are fitted with a combination of a Gaussian function and a linear function. 
In Figure~\ref{fig:moss_hist} we display both the $\sigma$ values of the Gaussian and the Full-Width-Half-Maximum 
($2\sqrt{2\ln(2)}\sigma\approx2.35\sigma$) values. The measured widths are comparable to the results obtained for coronal loops (e.g., \citealp{BROetal2013}) and for chromospheric structures (e.g., \citealp{DEPetal2007c}; \citealp{MORetal2012c}).

\subsection{Dynamics of the moss}
\begin{figure*}[!tp]
\centering
\includegraphics[scale=0.65, clip=true, viewport=0.0cm 0.0cm 27.cm 16.cm]{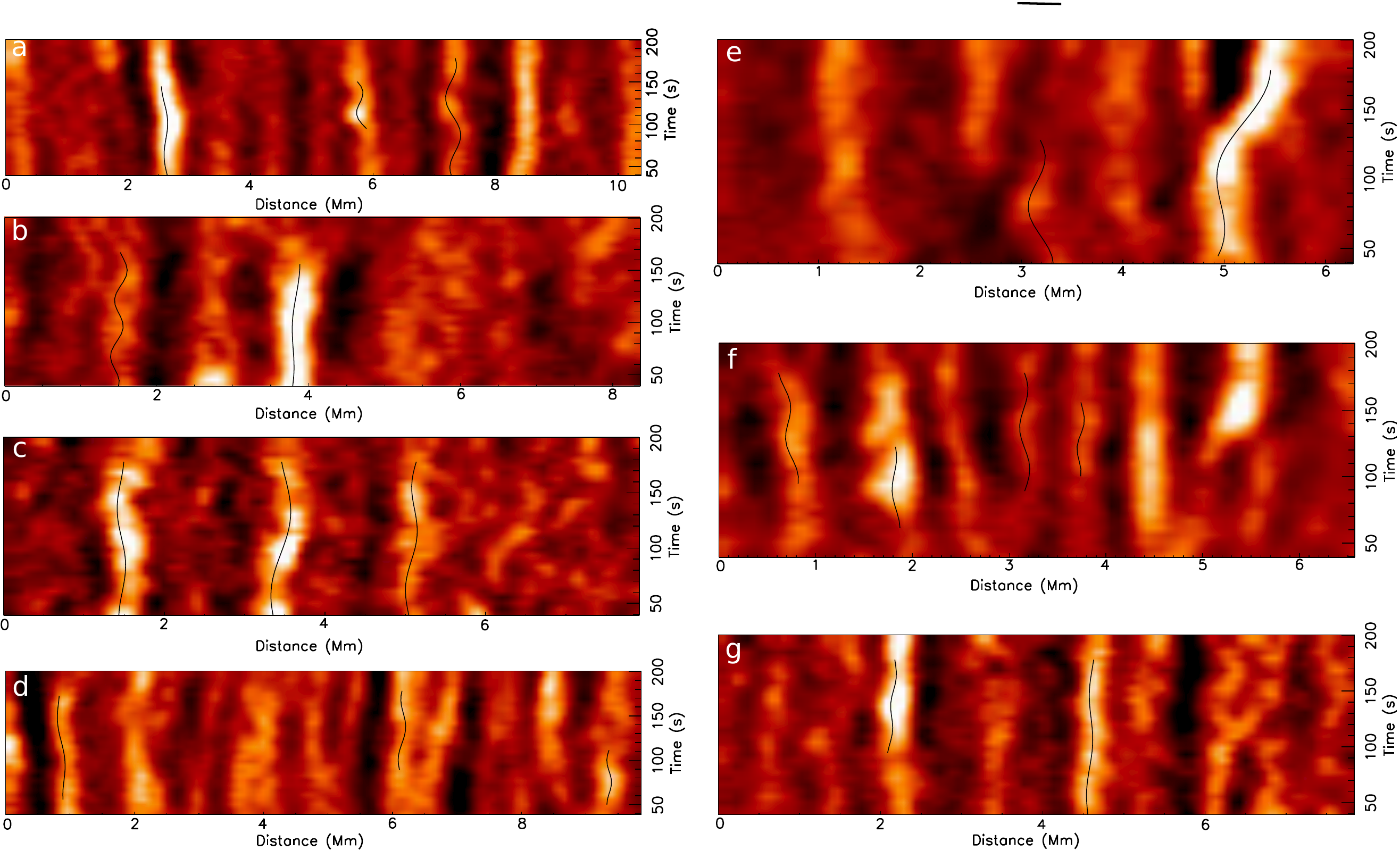}
\caption{Time-distance diagrams revealing {periodic transverse displacements} in active region moss features. {The plots are generated from the 
processed data.} Over plotted black lines are the results of sinusoidal fits to the observed displacement of the moss fine structure. The measured data 
points for each thread are shown in Figure~\ref{fig:oscill2} and details of the fit parameters are given in Table~1. Time is given in seconds from the start 
of the \textit{Hi-C} observations and the distance corresponds to the position along the cross-cut used.
}\label{fig:oscill}
\end{figure*}

Being able to resolve the fine structure now allows for the examination of the dynamic behaviour in the moss regions. {The data reveal that the 
fine-structure of the moss exhibits motions in the direction transverse to its axis, some of which demonstrate periodic behaviour. Examples of the 
observed {transverse displacements} are displayed in Figure~\ref{fig:oscill}.} 

General information about the techniques used in this paper to track and measure the {transverse displacements} are described in detail in 
\cite{MORMCL2013}. {However, we have advanced our analysis techniques and provide a brief description of them here.}

Due to the high read noise of \textit{Hi-C}, we {apply a filtering technique to each frame to suppress the the 
highest frequency spatial components. First, an Atrous filter is applied to each frame which extracts high frequency spatial components. The resulting 
high-frequency images still show signs of distinct structure. We then unsharp mask the high frequency component with a 3 by 3 boxcar function. This 
allows us to isolate a significant portion of the noise while minimising any potential signal loss. In theory, this procedure should separate the noise with a 
spatial variation less than 3 pixels. The diffraction limited seeing of Hi-C is $\sim0.3''$ (around 3 pixels - \citealp{KOBetal2014}); hence, the spatial 
variations are below the diffraction limit. The residual noise image shows a flat power spectral density and suggests uncorrelated noise (as also found in 
\citealp{KOBetal2014}). The noise image is then subtracted from the original data. This technique reduces the amount of signal loss 
compared to that used in \cite{MORMCL2013}, while still significantly improving the visibility of small-scale features in images by the removal of the 
majority of the read noise.}

The data is then subject to unsharp masking and cross-cuts are taken perpendicular to features of interest and time-distance diagrams
are created. {The time-distance diagrams are generated by averaging the intensities over two neighbouring cross-cuts.} The time-distance 
diagrams are then smoothed in space and time using a 3 by 3 pixel box-car function to suppress some of the additional large amplitude noise that arises 
from frame-to-frame variations in intensity levels. This aids the feature tracking routine that is then employed (e.g., \citealp{MORetal2013}), where a 
Gaussian function is fitted to the cross-sectional flux profile of each feature. The fit is supplied with the associated errors in data number, which are 
calculated using the formulae given in \cite{MORMCL2013} and {divided by a factor of $\sqrt{2}$ to account for the averaging over two 
neighbouring cross-cuts}.

{We focus on measuring transverse motions that display potential evidence for periodicity. The motions are fit with a combination of a sinusoidal 
function and a linear function (e.g. \citealp{MORMCL2013}). Upon testing the wave fitting routine on example data, the routine was able to detect 
periodic displacement amplitudes on the order of $0.2$~pixel (i.e. peak-to-peak displacement of $0.4$~pixel) for a signal-to-noise ratio of $\sim10$ or 
greater. For the fine-structure observed in the moss regions, the signal-to-noise is $\gtrsim20$.} When fitting the {transverse displacements}, we 
require a minimum of $3/4$ of a cycle for it to {be considered a potential signature of periodic behaviour. The selection of $3/4$ of a cycle is 
used since it ensures that the motion is observed at least in two directions and also allows for the peak-to-peak amplitude to be measured.}

From the time-distance diagrams it is possible to measure the transverse displacement amplitude ($\xi$) and the period ($P$) of the waves. The velocity 
amplitude, $v$, and its associated error can be calculated from these two quantities,
\begin{equation}
v=\frac{2\pi\xi}{P},\qquad \frac{dv^2}{4\pi^2}=\left(\frac{d\xi}{P}\right)^2+\left(\frac{\xi dP}{P^2}\right)^2,
\end{equation} 
where $dx$ is the error of the quantity $x$.

\begin{figure*}[!tp]
\centering
\includegraphics[scale=0.85, clip=true, viewport=0.0cm 0.0cm 20.cm 21.6cm]{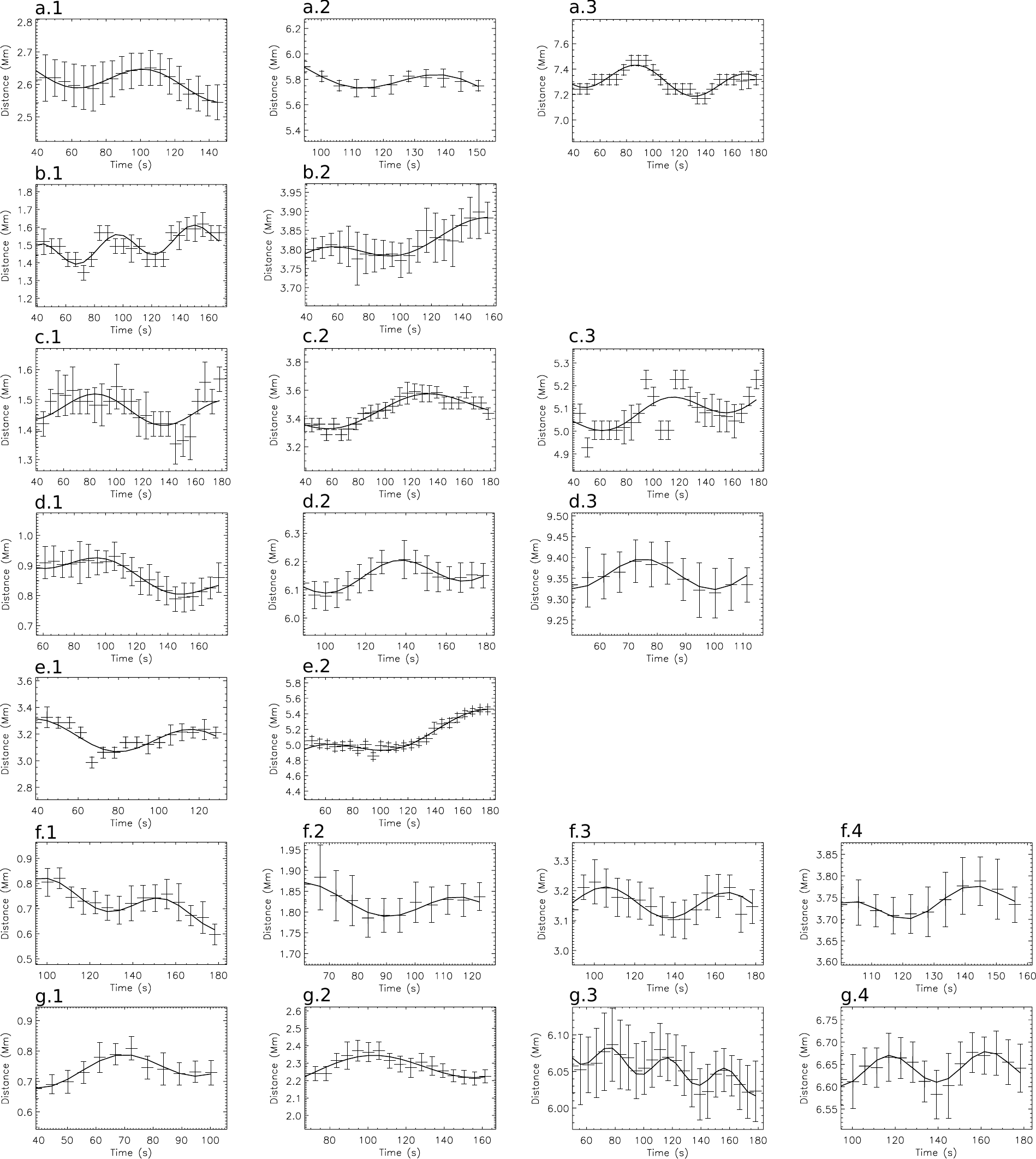}
\caption{Data points corresponding to highlighted threads in Figure~\ref{fig:oscill}. The given letter for each panel relates to the panel label in 
Figure~\ref{fig:oscill}. The $\sigma$ uncertainties on the position of each data point are given by the error bars. Over plotted black lines are the 
results of sinusoidal fits to the data points. The details of the fit parameters are given in Table~1. 
}\label{fig:oscill2}
\end{figure*}

In the identified moss regions, we find 73 measurable examples of {transverse displacements that demonstrate evidence for periodicity}. {A 
complete list of the measurements of $\xi$, $P$ and $v$ are given in Table~1 along with the duration of each signal. Typical examples of time-distance 
diagrams are shown in Figure~\ref{fig:oscill}, with the associated measured data points and sinusoidal fits given in Figure~\ref{fig:oscill2}. The majority 
of the measured transverse displacements show at least 
one cycle and typically only motions with periods longer than $\sim150$~s show less than a whole period.} Figure~\ref{fig:waves_hist} gives the 
histograms for the measured {transverse displacements}, which have means and standard deviations of $55\pm37$~km for the transverse 
displacement amplitude, $77\pm33$~s for the period and $4.7\pm2.5$~km/s for the velocity amplitude. {In addition to the periodic transverse 
displacements, some of the threads also display evidence for transverse motions that occur over longer time-scales. These motions cause the thread to 
appear to drift from its original position in the time distance diagram. This motion can also be superimposed with the periodic motions, as evidenced in 
e.g., Figure~\ref{fig:oscill2} b.2, e.2, g.3. }

\section{Discussion and conclusions}
Moss regions observed in EUV passbands are thought to be the high pressure Transition Region of hot coronal loops. The fine-scale structuring of cooler 
($T<$~2~MK) coronal loops has been evident from measurements in EUV images (e.g., \citealp{WATKLI2000}; \citealp{BROetal2012, BROetal2013}). 
Earlier observations from \textit{TRACE}, \textit{EIT} and \textit{Yohkoh} suggested that loop widths may increase with temperature (\citealp{SCH2007}), 
although this could be related to the lower spatial resolution of these previous missions. On the other hand, the observed width increase may be related 
to an increasing {\lq fuzziness\rq} of loops with increasing temperature (\citealp{BRISCH2006}; \citealp{TRIetal2009}). However, there is a suggestion 
that the observed fuzziness decreases for passbands sensitive to temperatures greater than 3~MK (\citealp{GUAetal2010}) and is apparently confirmed by 
\textit{SDO} observations (\citealp{REALetal2011}) although no measurements of individual loop widths are given. This would suggest that the hotter 
loops are structured on similar scales to the EUV loops.

\subsection{Spatial scales of moss features}
The increased resolution of \textit{Hi-C} allows the fine-scale magnetic structure to be resolved in moss regions. The fine-scale structure is most evident 
in the dark inclusions accompanying the bright moss emission. The patches of bright moss emission correspond to the Transition Region of a collection 
of closely packed loop legs and the close proximity of the loops obscure the lower-altitude portion of the loop leg. When neighbouring groups of 
loops have different inclinations, a gap appears in the moss (i.e., a dark inclusion) and elongated fine-scale structure, corresponding to the 
lower-altitude portion of the loop leg, becomes visible (e.g., Figure~\ref{fig:moss}). Similar structures are seen at the edges of moss regions when there are no neighbouring loops 
to obscure the view (Figure~\ref{fig:moss2}).  The resolved features have measured widths similar to fine-scale structure in the EUV corona 
(\citealp{BROetal2013}). 

It is interesting to note that the mean width ($\sigma\approx188$~km, FWHM$\approx440$~km) of the structures measured here is larger 
than those obtained for {dynamic fibrils (\citealp{DEPetal2007c} - measured total width - $340\pm160$~km)} and quiet Sun fibrils  
(\citealp{MORetal2012c} - $\sigma\approx150$~km, FHWM$\approx360$~km) and smaller than 
those measured for coronal loops with \textit{Hi-C} (\citealp{BROetal2013} - $\sigma\approx272$~km, FWHM$\approx640$~km), which hints at 
magnetic flux-tube expansion between the chromosphere and corona. {However, the given values are comparable within $\sigma$.} In 
addition, direct comparisons between the results from different instruments is 
complicated by a number of factors. Firstly, the lower spatial resolution of \textit{Hi-C}, compared to \textit{ROSA}, may influence the distribution of the 
flux tubes widths measured here. Secondly, the measured width is not equal to the actual diameter of the structure, 
with a complex relationship existing between measured and physical loop widths dependent upon instrument characteristics (\citealp{LOPetal2006}; 
\citealp{BROetal2012,BROetal2013}). Finally, for all cases, i.e., chromospheric, Transition Region, and coronal magnetic structures, measurements are 
from small sample numbers. The combination of these factors limit the conclusions that can be drawn at this time regarding expansion.

\subsection{Temperature regime of the fine-structure}
The fine-structure has less emission than the bright moss, indicating the features contain plasma that is cooler than the bright moss and 
likely constitute the lower Transition Region/upper chromospheric sections of the hot loops. {Let us consider the line intensity of coronal plasma,
$$
I=\int G(T,n)n^2\,dz,
$$
where $G(T,n)$ is the contribution function, $T$ the temperature, $n$ the electron density. As the fainter emission appears to form in the lower 
extension of the moss, it is expected that $n$ should be larger in the lower atmosphere (e.g. compare electron densities from the model P chromosphere 
of \citealp{FONetal1993} to measurements of moss densities - \citealp{TRIetal2010}). The increase in $n$ means that the contribution function must 
decrease if intensity is to decrease (as demonstrated in Figure~\ref{fig:moss3}), hence, the temperature of the plasma is decreasing. The intensity 
decreases at a steady rate (i.e. no sharp drop in intensity is observed) along the fine-structure from the bright moss, implying a similarly steady decrease 
in temperature. However, the plasma must be hot enough to contribute to the $193$~{\AA} passband so must have a temperature greater than 
$0.1$~MK}. The ability of \textit{Hi-C} to resolve the structure in the dark inclusions and measure their faint emission is due to the larger 
effective area of \textit{Hi-C} along with the increased resolution, when compared to the \textit{AIA} 193~{\AA} channel. 

\cite{DEPetal1999} noted that features in the dark inclusions in the moss can be relatively well correlated with absorption features observed in H$\alpha$, i.e., dynamic fibrils. 
\textit{Hi-C} reveals dark striations are present amongst the faint emission, which may be the signatures of chromospheric material reaching high into 
the lower corona and absorbing/blocking the hotter Transition Region emission (\citealp{DEPetal2009d}). {It may also be argued that the faint emission 
features in the dark inclusions are composed of cool, chromospheric plasma, which may be visible due to partial absorption of EUV radiation or scattered 
light. {This may be true for the lowest section of the features but seems unlikely for the upper sections given the gradual decrease in intensity 
along the structures. Additionally, it would appear unlikely that the majority of these features are the upper atmospheric extensions of dynamic fibrils. 
This is due to the clear morphological differences between the sparsely populated jets seen in H$\alpha$ (e.g., \citealp{DEPetal2007c}) and the relatively 
dense population of features we observe with \textit{Hi-C}.  }

\begin{figure*}[!tp]
\centering
\includegraphics[scale=0.67, clip=true, viewport=0.0cm 0.0cm 27.cm 10.cm]{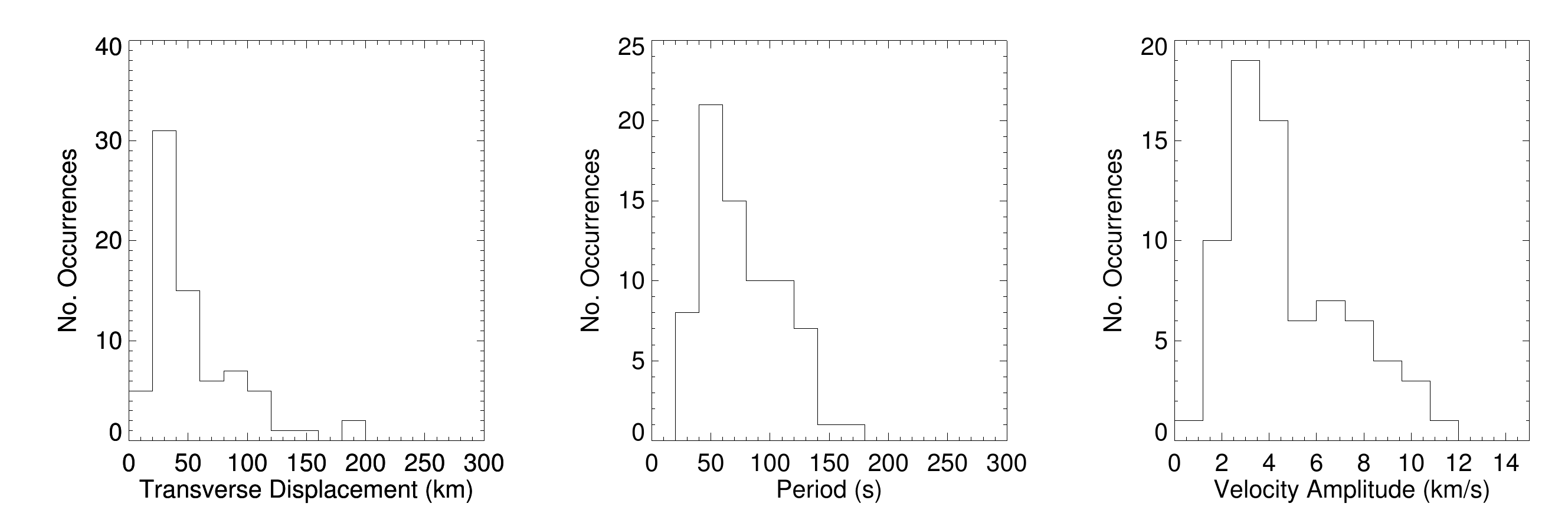} 
\caption{Histograms showing measured values of transverse displacement amplitude, period and velocity amplitude (from left to right) for
fine-scale Transition Region structures. The results have means and standard deviations of $55\pm37$~km for the transverse 
displacement amplitude, $77\pm33$~s for the period and $4.7\pm2.5$~km/s for the velocity amplitude.
}\label{fig:waves_hist}
\end{figure*}

\subsection{Transverse motions - signatures of kink waves}
The \textit{Hi-C} data also reveals the {transverse displacement} of the moss fine structure. The observed displacements
are of the fine-scale faint emission features that lie within the dark inclusions (see, Figures~\ref{fig:moss} and 
\ref{fig:mossa}). The ability to measure {transverse displacements} in the dark inclusions is limited in part due to the low S/N levels in these 
regions and partially due to the small displacements ($\sim50$~km) the features undergo. This latter point is clear when considering the relative errors 
obtained for the smallest displacements in Table~1. 

{Due to some of the transverse motions displaying evidence for periodic behaviour, we interpret the observed motions as the kink (Alfv\'enic) 
wave.} The presence of waves in the moss has been conjectured (\citealp{ANTetal2003}; \citealp{BROWAR2009}) and it is thought that the waves are the 
cause of some of the observed variability in the cores of the active region moss rather than heating events. {We recall here that \cite{BERetal1999} 
noted they observed the interaction of dynamic fibrils with the EUV moss, which occasionally pushed the moss elements aside. However, we suggest that 
the periodicity demonstrated by the motions reported here is a key indicator of wave behaviour, rather than the interaction of the dynamic fibrils with the 
EUV moss. Additionally, we do not see evidence for this interaction in the \textit{Hi-C} data, although its likely that further high resolution coronal 
observations with co-temporal chromospheric data are required to observe such an interaction.}

\subsection{Insights into wave propagation through the atmosphere}
The observation of the periodic transverse displacements in the moss features may suggests that kink (Alfv\'enic) waves can propagate into the 
Transition Region from the chromosphere. There is still an open question related to the fraction of Alfv\'enic wave energy, generated in the lower solar 
atmosphere, that is able to reach the corona. It has been suggested that a significant fraction of the energy is reflected at the Transition Region due to 
steep density gradients, with evidence from quiet Sun chromospheric observations suggesting around $40\%$ of the waves are reflected 
(\citealp{OKADEP2011}; \citealp{KURetal2013}). The high pressure nature of the hot loops may reduce the sharpness of density gradient in the Transition 
Region {(see, e.g. simulations of hot loops - \citealp{SERetal1981}; \citealp{REA2010})} and allow waves to propagate more freely from the 
chromosphere to the corona. 

{If we assume that the observed transverse displacements can be interpreted in terms of kink waves}, an estimate of the difference in energy, $E$, 
and Poynting flux, $S_z$, between the chromospheric and Transition Region waves can be made. To make this estimate we use 
the formulae for integrated energy and Poynting flux from \cite{GOOetal2013}. {In deriving these formulae, \cite{GOOetal2013} assume the 
wave-guide is a cylindrical, pressure-less, over dense flux tube. These assumptions are applicable to the fine-scale features we observe in the 
\textit{Hi-C} images and also to dynamic fibrils in the chromosphere. Both features are likely low-$\beta$ plasmas, where $c_s<<v_A$, hence, can be 
considered pressure-less analytically. Additionally, the kink mode is known to be highly incompressible in the long wavelength limit 
(\citealp{GOOetal2009}), so neglecting pressure will have little influence on the wave energy calculation. 
Further, the fine-scale features seen in \textit{Hi-C} appear to be the lower atmospheric extension of the bright moss, hence, they will also have a high 
pressure and a greater density than the ambient coronal plasma in which they reside. Dynamic fibrils are jets of chromospheric plasma into the upper 
atmosphere, so are also over dense compared to the ambient plasma.}  

The ratio of Transition Region to chromospheric wave energy and Poynting flux is then
\begin{eqnarray}
\frac{E_{TR}}{E_{c}}&\approx &\frac{\rho_{TR}v_{TR}^2R_{TR}^2}{\rho_{c}v_{c}^2R_{c}^2}=0.15,\\
\frac{S_{z,TR}}
{S_{z,c}}&\approx &\frac{c_{k,TR}}{c_{k,c}}\frac{E_{TR}}{E_{c}}=0.46.
\end{eqnarray}
Here, $\rho$ is the density and $R$ is the radius of the flux tube. The subscript TR and c correspond to Transition Region and chromosphere, 
respectively. The given ratios will likely be best described using frequency dependent functions. This is because damping mechanisms and reflection alter 
the wave amplitude and tend to show frequency dependent behaviour (e.g., \citealp{VERTHetal2010}, \citealp{MORetal2013b}). At present we have to 
ignore such complications due to a lack of information and have used the mean value of the velocity amplitude measured here, $v_{TR}=4.7$~km/s. {To 
the best of our knowledge, 
there are no measurements of transverse displacements in dynamic fibrils to provide direct comparison too. However, the measured properties of the 
transverse displacements are similar to those measured in active region fibrils ($4.4\pm2.3$~km/s; \citealp{MORetal2013b}), while the values are 
smaller than those reported for active region spicules (type-I - $\sim10$~km/s; 
\citealp{PERetal2012}).  We choose to use the values from active region fibrils ($v_c=4.4$~km/s), to provide an upper limit for the energy transmission.
It is clear that using active region spicules values of velocity amplitude ($\sim10$~km/s) would lead to much smaller ratios.}

 In addition, we have used $\rho_{TR}=10^{10}$~cm$^{-3}$  (e.g. \citealp{TRIetal2010}; 
\citealp{WINetal2011}), $\rho_c=10^{11}$~cm$^{-3}$ (\citealp{FONetal1993} - Model P), 
$R_{TR}=440$~km, $R_c=340$~km (\citealp{DEPetal2007c}) and characteristic values of the propagation speed of waves in the chromosphere 
($\sim100$~km\,s$^{-1}$ - \citealp{OKADEP2011}, \citealp{MORetal2012c}) and Transition Region ($\sim250$~km\,s$^{-1}$ - \citealp{MCIetal2011}) are used. 

\bigskip

The estimates reveal that the Transition Region Poynting flux carried by waves is potentially $40\%$ smaller than that in the 
chromosphere. However, the overall wave energy has decreased by $\sim85\%$.

Moreover, the measured velocity amplitudes of the waves here are greater than those typically reported in observations of coronal kink waves in active 
regions ($<2$~km/s - \citealp{TOMetal2007}; \citealp{ERDTAR2008}; \citealp{VANetal2008c}; \citealp{TIAetal2012}; \citealp{MORMCL2013}), which 
implies that kink waves are more energetic in the lower solar atmosphere compared to the active corona. {\cite{MCIetal2011} and 
\cite{MORMCL2013} both present evidence for larger velocity amplitude ($> 5$~km/s) transverse motions in the corona (unrelated 
to excitation by flare blast waves). However, as demonstrated in \cite{MORMCL2013}, it does not appear as if these larger amplitude motions are 
ubiquitous throughout the active corona. We suggested that observed transverse displacements with large velocity amplitudes are related 
to significant energy releases in individual loop systems, e.g. via magnetic reconnection, and not continuously driven.} 

These estimates suggest that the majority of observed kink wave energy in the chromosphere is unable to reach the corona, however, they do not rule 
out waves as contributing to the heating budget of hot loops. {The estimates suggest if waves are to play a role in active 
region heating then the energy deposition is likely to be localised in the chromosphere. Alternatively, the kink wave energy may have been converted to 
torsional motions of the fine-structure through mode-coupling via resonant absorption (e.g., \citealp{TERetal2010c}; \citealp{PASetal2011}). Similar 
findings are found for waves in quiet regions in \cite{MORetal2013b}.}  The continuous driving of waves via granular or turbulent motions (e.g., 
\citealp{VANBALLetal2011}) could provide the required quasi-steady heating mechanism in active regions. On the other hand, the observed waves could be the 
by-product of small-scale magnetic reconnection events, i.e., the classic nano-flares. It is well known from simulations of large-scale reconnection 
events that a fraction of the 
energy released is converted to wave energy (\citealp{YOKSHI1996}). The results presented here are a preliminary investigation into waves in the 
Transition Region but they point towards the need for extended statistical studies of wave propagation from the chromosphere to the corona. Such further study will be required to provide a clear and unequivocal picture of wave propagation through the atmosphere.

\begin{acknowledgements}
The Authors thank the referee whose comments have helped improved the manuscript. RM is grateful to Northumbria University for the award of the 
Anniversary Fellowship and thanks D. Brooks and G. Verth for 
useful discussions. The authors acknowledge IDL support provided by STFC. JM contribution forms part of the effort sponsored by the Air Force Office of 
Scientific Research, Air Force Material Command, USAF, under grant number FA8655-13-1-3067. The U.S Government is authorized to reproduce and 
distribute reprints for Governmental purpose notwithstanding any copyright notation thereon. We 
acknowledge the High resolution Coronal imager instrument team for making the flight data publicly available. 
MSFC/NASA led the mission and partners include the Smithsonian Astrophysical Observatory in Cambridge, Mass; 
Lockheed Martin's Solar Astrophysical Laboratory in Palo Alto, Calif; the University of Central Lancashire in 
Lancashire, UK; and the Lebedev Physical Institute of the Russian Academy of Sciences in Moscow.
\end{acknowledgements}

\begin{center}
\begin{deluxetable}{cccc}\label{tab:meas}
\tablewidth{200pt}

\tablecaption{Measured properties of the displacements}\label{tab:A1}

\tablehead{\colhead{$\xi$ (km)} & \colhead{$P$ (s)} & \colhead{$v$ (km/s)} & \colhead{Signal Length (s)}}

\startdata

$      45\pm      11$&$      97\pm      11$&$2.9\pm0.78$&     145\\
$      19\pm      10$&$      33\pm       5$&$3.7\pm2.1$&      72\\
$      35\pm      14$&$      48\pm       5$&$4.6\pm1.9$&     100\\
$      16\pm       6$&$     154\pm      52$&$0.7\pm0.3$&     145\\
$      18\pm      11$&$      55\pm      11$&$2.0\pm1.3$&     111\\
$      24\pm      12$&$      50\pm      10$&$3.0\pm1.6$&      84\\
$     100\pm      13$&$      97\pm       5$&$6.5\pm0.9$&     145\\
\tablenotemark{f.1} $      45\pm      19$&$      53\pm       8$&$5.3\pm2.3$&      89\\
\tablenotemark{f.2} $      32\pm      18$&$      55\pm      21$&$3.7\pm2.4$&      67\\
\tablenotemark{f.3} $      48\pm      21$&$      60\pm       7$&$5.0\pm2.3$&      95\\
\tablenotemark{f.4} $      29\pm      22$&$      39\pm      12$&$4.6\pm3.7$&      61\\
$      72\pm      50$&$     107\pm      27$&$4.2\pm3.1$&     100\\
$      29\pm      22$&$      42\pm       7$&$4.3\pm3.3$&      78\\
$      69\pm      29$&$     113\pm      33$&$3.8\pm2.0$&     123\\
$      39\pm      21$&$     119\pm      41$&$2.0\pm1.3$&     145\\
$      45\pm      28$&$      40\pm      12$&$7.0\pm4.8$&      56\\
$      37\pm      21$&$     125\pm      61$&$1.9\pm1.4$&     123\\
$      24\pm      20$&$      46\pm      18$&$3.3\pm3.0$&      72\\
\tablenotemark{a.1} $      40\pm      17$&$      85\pm      21$&$2.9\pm1.4$&     111\\
\tablenotemark{a.2} $      97\pm      87$&$      59\pm      35$&$10.0\pm11.0$&      61\\
\tablenotemark{a.3} $     106\pm      12$&$      83\pm       3$&$8.0\pm1.0$&     145\\
$      49\pm      18$&$      35\pm       3$&$8.7\pm3.3$&      78\\
$      59\pm      16$&$      49\pm       4$&$7.6\pm2.1$&     100\\
$      32\pm      21$&$      40\pm       5$&$5.0\pm3.4$&      84\\
$      32\pm      15$&$      61\pm      10$&$3.3\pm1.7$&     106\\
$      36\pm      15$&$      79\pm       9$&$2.9\pm1.2$&     145\\
$      29\pm      19$&$      57\pm      14$&$3.2\pm2.2$&      84\\
$      72\pm      19$&$      92\pm      18$&$4.9\pm1.6$&     106\\
$      66\pm      18$&$      63\pm      13$&$6.5\pm2.2$&      72\\
$      56\pm      25$&$     125\pm      39$&$2.8\pm1.5$&     128\\
$      35\pm      16$&$      67\pm      11$&$3.2\pm1.6$&     106\\
$      25\pm      17$&$      38\pm       5$&$4.0\pm2.8$&     117\\
$      31\pm      10$&$     114\pm      23$&$1.7\pm0.7$&     145\\
$     151\pm      84$&$     132\pm      46$&$7.2\pm4.7$&     106\\
$      31\pm      27$&$     124\pm      35$&$1.6\pm1.5$&     134\\
$      33\pm      17$&$      49\pm      11$&$4.3\pm2.4$&      67\\
$     189\pm     581$&$     372\pm     464$&$3.2\pm11.0$&     145\\
$      19\pm      14$&$      92\pm      25$&$1.3\pm1.0$&     117\\
$      82\pm      48$&$     122\pm      49$&$4.2\pm3.0$&     134\\
$      31\pm      19$&$      45\pm       9$&$4.3\pm2.9$&      72\\
$      28\pm       8$&$      73\pm       7$&$2.4\pm0.8$&     145\\
\tablenotemark{e.1} $     103\pm      15$&$      78\pm       7$&$8.4\pm1.5$&      95\\
\tablenotemark{e.2} $     133\pm      12$&$     117\pm       8$&$7.1\pm0.82$&     139\\
$      22\pm      16$&$      75\pm      21$&$1.9\pm1.4$&     111\\
$      64\pm      21$&$      44\pm       4$&$9.3\pm3.1$&      72\\
$      54\pm      19$&$     116\pm      32$&$3.0\pm1.3$&     145\\
$      45\pm      19$&$     113\pm      39$&$2.5\pm1.4$&     145\\
$      49\pm      19$&$      40\pm       3$&$7.7\pm3.1$&      84\\
$      34\pm      22$&$      31\pm       7$&$6.9\pm4.8$&      50\\
$      29\pm      15$&$      90\pm      13$&$2.0\pm1.1$&     145\\
$      23\pm      16$&$      48\pm      10$&$3.1\pm2.2$&      89\\
$      98\pm      19$&$      62\pm       4$&$10.0\pm2.0$&     100\\
$      97\pm      19$&$      68\pm      10$&$9.0\pm2.2$&      95\\
$      81\pm      18$&$      78\pm      11$&$6.5\pm1.7$&     111\\
$     182\pm      14$&$     127\pm       8$&$9.0\pm0.9$&     145\\
$      32\pm      18$&$      37\pm       4$&$5.5\pm3.2$&      89\\
$      53\pm      12$&$     103\pm      10$&$3.2\pm0.8$&     145\\
$      22\pm      11$&$      55\pm       8$&$2.5\pm1.4$&      95\\
$      28\pm      15$&$      63\pm      10$&$2.8\pm1.5$&     100\\
\tablenotemark{c.1}$      47\pm      13$&$      99\pm      20$&$3.0\pm1.0$&     145\\
\tablenotemark{c.2}$      94\pm      20$&$     136\pm      24$&$4.3\pm1.2$&     145\\
\tablenotemark{c.3}$      53\pm      13$&$      94\pm       7$&$3.5\pm0.9$&     145\\
$     109\pm      15$&$      71\pm       7$&$9.6\pm1.7$&      84\\
$     111\pm      61$&$      63\pm      26$&$11.\pm7.6$&      56\\
\tablenotemark{g.1} $      45\pm      15$&$      55\pm      11$&$5.1\pm1.9$&      67\\
\tablenotemark{g.2} $      81\pm      68$&$     115\pm      58$&$4.4\pm4.3$&     100\\
\tablenotemark{g.3} $      15\pm      10$&$      40\pm       5$&$2.4\pm1.6$&     134\\
\tablenotemark{g.4} $      32\pm      17$&$      45\pm       7$&$4.5\pm2.5$&      89\\
$     183\pm      20$&$     165\pm      19$&$7.0\pm1.1$&     145\\
\tablenotemark{b.1} $      70\pm      13$&$      53\pm       3$&$8.3\pm1.6$&     134\\
\tablenotemark{b.2} $      29\pm      12$&$     100\pm      26$&$1.8\pm0.9$&     123\\
\tablenotemark{d.1} $      36\pm      14$&$      87\pm      23$&$2.6\pm1.2$&     123\\
\tablenotemark{d.2} $      47\pm      18$&$      68\pm       9$&$4.3\pm1.7$&      95\\
\tablenotemark{d.3} $      36\pm      22$&$      49\pm      14$&$4.7\pm3.1$&      67\\

\enddata
\tablenotetext{a.1-g.4}{Notes give references to periodic motions shown in Figures~\ref{fig:oscill} and \ref{fig:oscill2}}

\end{deluxetable}
\end{center}

\end{document}